\pdfoutput=1
\documentclass [twocolumn,pre]{revtex4}   

\usepackage[ansinew]{inputenc}
\usepackage[T1]{fontenc}
\usepackage{graphicx,textcomp,booktabs,amsmath}
\usepackage{mathptmx,courier}
\usepackage{txfonts}
\usepackage[english]{babel} 
\usepackage{color}
\usepackage{tabularx}
\usepackage[FIGTOPCAP, raggedright]{subfigure}
\usepackage[a4paper,left=1.5cm, right=1.5cm, top=1.5cm, bottom=2.1cm]{geometry}

\usepackage{microtype}

\usepackage{nicefrac}
\newcommand{\SES}{state of enhanced activity }
\newcommand{\SESs}{states of enhanced activity }

\newcommand{\PDF}{probability density function }

\begin{document}

\title{States of Enhanced Activity in a Network of 
Pulse Coupled Oscillators with Dynamic Coupling}

\author{Daniel Ritterskamp}
\author{Rudolf Friedrich}
\affiliation{Institute for Theoretical Physics, University M\"unster, 48149 M\"unster, Germany}
\date{\today}%

\begin{abstract}
\noindent We investigate \SESs in a biological neuronal network
composed of pulse-coupled oscillators.
The synaptic couplings between the neurons are dynamic, modeling
spike time dependent plasticity. 
The network exhibits 
statistical characteristics which recently have been 
identified in an analysis of
epileptic seizures [Osorio et al., Phys. Rev. E{\bf 82}, 021919(2010)] 
based on analogies to the onset of earth quakes. 
\end{abstract}

\maketitle

\section{Introduction}

\noindent The origin of epileptic seizures is still a mystery 
(see e.g. \cite{Browne2008book}). 
The epileptic brain exhibits spatio-temporal complexity, both
during seizures as well as in the inter-seizure intervals. 
In recent years, the dynamical 
properties of inter-seizure states have been in the focus of research with
the aim to formulate suitable measures for the prediction of seizure 
onset \cite{Elger1998eujouneu}.   

A major goal of theoretical seizure modeling is the 
development of models of neuronal activity, which are able to reproduce 
the observations and allow for a theoretical interpretation 
of the spatio-temporal behaviour
of seizure states. A successful model will shed considerable light on the
questions whether seizures are predictable and what measures can be rewarding. 
The basic issues of theoretical 
seizure modeling are to investigate how epileptic 
seizures emerge, spread, and 
terminate in such models. Dynamical systems theory points to the 
importance of synchronization, however,
it seems to be rather unclear whether the 
synchronization of neuronal activity is a by-product, whether it 
induces  or whether it terminates a seizure 
\cite{Dominguez2005joneuro, Lehnertz2009joneuromethod}.

Recently, Osorio et al. \cite{Osorio2010pre} opened up a new perspective
for our understanding of epilepsy
assessing statistical similarities between 
epileptic seizures and the onset of 
earthquakes. Based on a comparative study of seizure and earth quake data, 
the authors showed ample evidence that the Gutenberg-Richter law, 
the Omori law and the statistics of interevent times formulated for 
earth quake data can be detected in epileptic seizure data as well. 

The purpose of the present article is to develop a neural network model, 
which is able to reproduce
the main characteristics of the analysis of Osorio et al. 
\cite{Osorio2010pre}. The developed model is based on a 
generalization of Haken's Lighthouse model (for a review we refer the reader 
to \cite{Haken2006book}), which is augmented by a suitable learning 
algorithm taking into account
the phenomenon of spike-timing-dependend plasticity
(STDP), following closely the work of Chen et al. \cite{Chen2010pre}. 
The model presented in this article
can be viewed as a model for epileptic seizures
similar to the Burridge-Knoppoff model
\cite{Burridge-Knopoff1967} and its variants for earth quakes.
Our extension of the Lighthouse model allows for the inclusion of a 
spontaneous restructuring of the network architecture, making use of the 
idea that dynamical seizures of the brain can be learned or dislearned 
\cite{Tass2008book}, \cite{Hsu2012}.

It is well-known that adaptive networks may lead to the emergence of
critical states in neural network models based on
the phenomenon of self-organized criticality
(SOC), as described by P. Bak \cite{Bak1999book}.
Experimentally,  
Beggs and Plenz \cite{BeggsJoneu2003} showed convincing evidence that the 
dynamics of neuronal populations could actually exhibit critical dynamics
by an explicit investigation of events called neuronal avalanches.
Neuronal avalanches can be found in regions of control parameter space
located in between a regime of coherent, wavelike neuronal activity and 
a regime, where neurons behave asynchronous and spike incoherently. 
The existence of SOC 
in an integrate and fire model including dynamical synapses 
has been shown by A. Levina et al. \cite{Levina2007, Levina2009prl}. 
The dynamics of their model exhibits the occurence of critical avalanches 
and the distribution of the size of the
occurred avalanches showed a power law behaviour. 
In the case of static synapses, the 
coupling strengths have to be fine-tuned in order to generate a 
critical state, whereas dynamical synapses renders the system critical. 
Similar results were obtained by L. de Arcangelis et al. \cite{Arcangelis2006}.
C. Meisel et al. \cite{Meisel2009pre} presented a study of neural networks including spike-time-dependent synaptic plasticity, extending the earlier work of 
Bornholdt and Rohlf \cite{Bornhioldt2003pre}. They showed that due to the 
inclusion of this learning mechanism the network self-organizes into a 
critical state. Millman et al. \cite{Millman2010} demonstrated the 
existence of SOC in nonconservative models of
networks of leaky integrate and fire neurons with short-term synaptic 
depression. 
They demonstrated the existence of two states, up and down states and demonstrated that up states are
critical, whereas down states are subcritical. 

With respect to epileptic seizures it is not clear whether SOC plays
a dominant role. Recently, it has been discussed that adaptive self-organized
criticality fails during epileptic seizure attacks \cite{Gross2012}

\noindent 
The present article is outlined as follows.
In the first section, we review Haken's Lighthouse model and discuss its 
extension taking into account a learning mechanism based on the mechanism
of spike-timing-dependent plasticity. 
In the second section, we investigate in detail states of enhanced activity.
We reproduce features similar to the ones
presented in \cite{Levina2009prl, Rothkegel2011epl}.
Then, we focus on similarities with the investigations of
experimental recordings of seizure
states discussed in \cite{Osorio2010pre}.
A central issue is the question why \SESs
emerge and terminate in the used model. 

\section{Model and Implementation}

\subsection{Lighthouse Model }

The Lighthouse model has been introduced by Herman Haken and is described
in great detail in his monograph \cite{Haken2006book}.
It models brain tissue as a neuronal network 
composed of pulse coupled oscillators. 
Each neuronal oscillator, labeled by the index $m$, is characterized 
by its phase $\phi_m$, a $2 \pi$- periodic variable,
and a dendritic current $\psi_m$.  \\
\noindent The phase describes the action potential of a neuron. If the phase 
reaches the threshold $2 \pi$, a neuron will spike.
The action potential $S_k$ generated by neuron $k$ is obtained as a 
sum over delta functions:
    \begin{align}
	S_k(\phi_k(t)) = \sum_n \delta(2\pi - \phi_k(t_n)) ~ \dot \phi_k(t)
	\label{eqn: Spike_function}
    \end{align}
The temporal evolution of the phase is determined by the equation 
    \begin{align}
      \label{eqn: Phase}
      \dot{\phi}_m(t) = \Xi \left(\sum_{k} c_{mk} \psi_{k}(t) + p_{(ext,m)}(t) , \Theta \right) 
    \end{align}
where $\Xi \left(\sum_{k} c_{mk} \psi_{k}(t) + p_{(ext,m)}(t) , \Theta \right)$ 
is a sigmoidal function of the sum of all dendritic currents $\psi_k$ of
neurons $k$ mutiplied by a weighting factor $c_{mk}$ and an external 
signal $p_{(ext,m)}(t)$. 
The factors $c_{mk}$ form the increment matrix $\bar{c}$.
In the following, we use a diagonal increment matrix $c_{mk}=c_m \delta_{mk}$.
The sigmoidal function $\Xi(X, \Theta)$ is taken as 
the Naka-Rushton relation and describes 
the response of a neuron to an input current. 
    \begin{align}
	\Xi(X,\Theta)=\frac{\nu X^M}{\Theta^M+X^M}
	\label{eqn: Naka-Rushton relation}	
    \end{align}
For high input values $X$ the output saturates at a maximal 
firing rate $\nu$. If the input signal is below a certain
threshold value $\Theta$, the output will tend to zero. 
As outlined in \cite{Haken2006book} this models the 
all-or-none-behaviour of the axon hill, which generates the action potential.
Due to the saturation of the Naka-Rushton relation for large input values, a 
refractory period of the neurons is included in the system.
The constant $M$ is related to the slope of the function.
In our treatment we have used $M=3$ and a maximal firing rate $\nu$ of $1$ .

When a neuron generates an action potential, this will be transfered to
all connected neurons. The dendritic current $\psi_m(t)$ of neuron 
$m$ will change due to the
input from neuron $k$. It obeys the following differential equation
    \begin{align}
	\dot{\psi}_m(t) = \sum_{k} a_{mk} ~ S_k(\phi_k(t)) - \gamma \psi_m(t).
	\label{eqn: Dendritic current}
    \end{align}	
The synaptic weights are denoted by the matrix elements $a_{mk}$ and describe the
strength of the connection from neuron $k$ to neuron $m$.
The synaptic weights characterize the connections of the 
neural network. 
In the following self-coupling is neglected, i.e. $a_{mm}=0$.

\subsection{Spike-Timing-Dependent Plasticity }

In order to include learning, we combine the Lighthouse model 
with the mechanism of spike time dependent 
plasticity. For a historical review we refer the reader to
\cite{Markrsam}. This mechanism is 
based on the Hebbian learning rule. We will rely on the
formulations of C. Chen and 
D. Jasnow \cite{Chen2010pre} and van Rossum et al. \cite{Rossum2000JoN}. 
The underlying mechanism of STDP is as follows: \\
The coupling weight $a_{km}$ will be 
strengthened due to causal firing and weakened in the case of acausal firing. 
Causal firing means that neuron $m$ spikes in advance of neuron $k$. 
Acausal firing 
denotes the reverse case. The reverse coupling weight $a_{mk}$ will be weakened 
if $a_{km}$ is strengthened and vice versa. Hence, 
the connections between two neurons become directed and dynamical.
 
In order to formulate a mathematical 
model of spike time dependent plasticity one has to design a mechanism, 
which is able to distinguish causal from acausal spiking. 
Spiking events of neuron $i$ are
encoded in the spike train $S_i=\sum_{n}\delta(t-t_i^n)$. 
To the $n$-th firing event of neuron $j$ at time $t_j^n$ we connect a function 
$\sigma_j(t)=\sigma_0 \Theta (t-t_j) e^{-\frac{(t-t_j^n)}{\tau_{\sigma}}}$.
The product $\sigma_j(t)~ S_i(t)$ is 
zero if $t_i^n<t_j^n$, whereas it is nonzero in the reverse case.
We have to assume that the decay time $\tau_{\sigma}$ is smaller than the
characteristic inter-spike intervals $t_{j+1}-t_j$. 
The coupling weight $a_{ij}$ decreases in the first (acausal) case.
A similar reasoning can be used for the reversed cased. 
It is convenient to introduce
two functions $\sigma_j(t)$, 
denoted as $A_j(t)$ and $B_j(t)$, for each
neuron $j$.

The variable $A_m(t)$ associated to neuron $m$
is related to strengthening of the connection
$a_{km}$ if the firing with respect to neuron $k$ is causal, i.e. if the
product $A_m(t)S_k(t)$ is different from zero. The variable
$B_k(t)$ induces weakening of the coupling strength $a_{km}$ if the firing
with respect to neuron $m$ is acausal, i.e if the product $B_k(t)S_m(t)$ 
is different from zero. The dynamics of the coupling weight is then given
by the evolution equation  
	\begin{align}
		\dot{a}_{km} = \Delta A_m S_k - r a_{km} B_k S_m
		\label{eqn: dynamic synaptic strength}
	\end{align}
The first term of equation (\ref{eqn: dynamic synaptic strength}) 
is related to strengthening due to causal firing and the amount is proportional 
to the potentiation constant $\Delta$. 
The second term causes depression of the coupling weight. 
It is proportional to the depression constant $r$ and
to the coupling weight $a_{km}$ itself. 

In order to determine the functions $A_m(t)$ and $B_m(t)$ dynamically we closely follow
the work of Chen and Jasnow \cite{Chen2010pre}. They used 
the following evolution equations, where $\sigma_m(t)$ now denotes
$A_m$ and $B_m$. The quantities $A_m$ and $B_m$ can be viewed as concentrations of
chemical species, which are generated by a spiking event of the neuron $m$.  

\begin{align}
		\dot \sigma_m = u_{\sigma}(1-\sigma_m- I_{\sigma}) S_m 
- \frac{\sigma_m}{\tau_{\sigma}} ~~~; ~~~ \sigma = A,B
		\label{eqn:Chemical concentrations}
	\end{align}
Furthermore, the dynamics of $I_\sigma$ is given by
\begin{align}
 \dot I_{\sigma} = \frac{\sigma}{\tau_{l\sigma}} - \frac{I_{\sigma}}{\tau_{R\sigma}}
 \label{eqn: fatigue}
\end{align}
The spiking of a neuron, encoded in $S_m$, leads to an increase of the
concentrations $A_m$, $B_m$. Furthermore, the concentrations decay with
time constants $\tau_\sigma$. For details we refer the reader to
\cite{Chen2010pre}.

\begin{figure}[t]
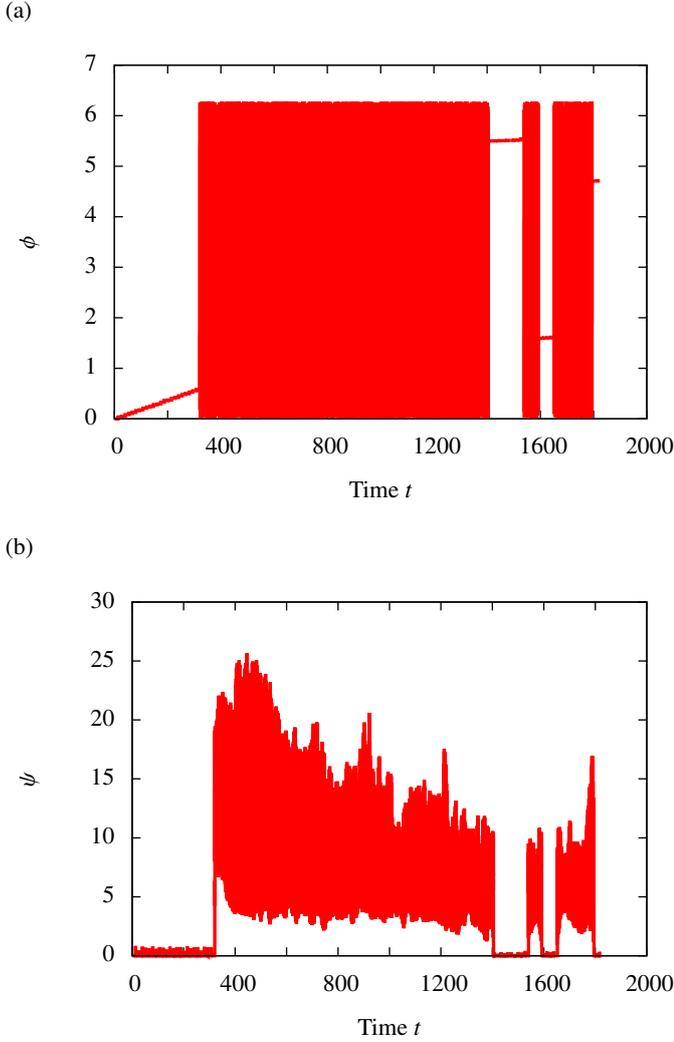

  \begin{tabular}{l}
   (a) \\
	 \scalebox{1}{\input{Phase_Paper} }  \\
   (b) \\
	 \scalebox{1}{\input{Current_Paper} }  \\   
  \end{tabular}

  \caption{(a) Phase $\phi$ of a single neuron is shown as a function of time. 
(b) The dendritic current $\psi$ of the same neuron is plotted in time. 
Figures (a) and (b) demonstrate the existence 
of time-intervals, during which the neuron is at rest and intervals when it is higly active. }
  \label{fig: SLS_CurrentPhase}
\end{figure}

\section{States of Enhanced Activity}

\noindent  The aim of this article is to find states in the extended
Lighthouse model, which exhibit features of epileptic seizures. 
There seems to be no commonly accepted definition of a seizure in literature.
However, as a main feature one may consider 
the fact that all neurons in a region with epileptic spiking are  
active and spike with a high frequency.
Synchronization in the sense of dynamical systems theory seems to
be an acompanying phenomenon, as pointed out by R. S. Fisher 
et al \cite{Fisher2005epi}. However, whether synchronization is the origin of 
a seizure or plays a major role in the terminiation, as emphasized by
\cite{Lehnertz2009joneuromethod, Schindler2008chaos}, is yet unclear.

\begin{figure}[t]
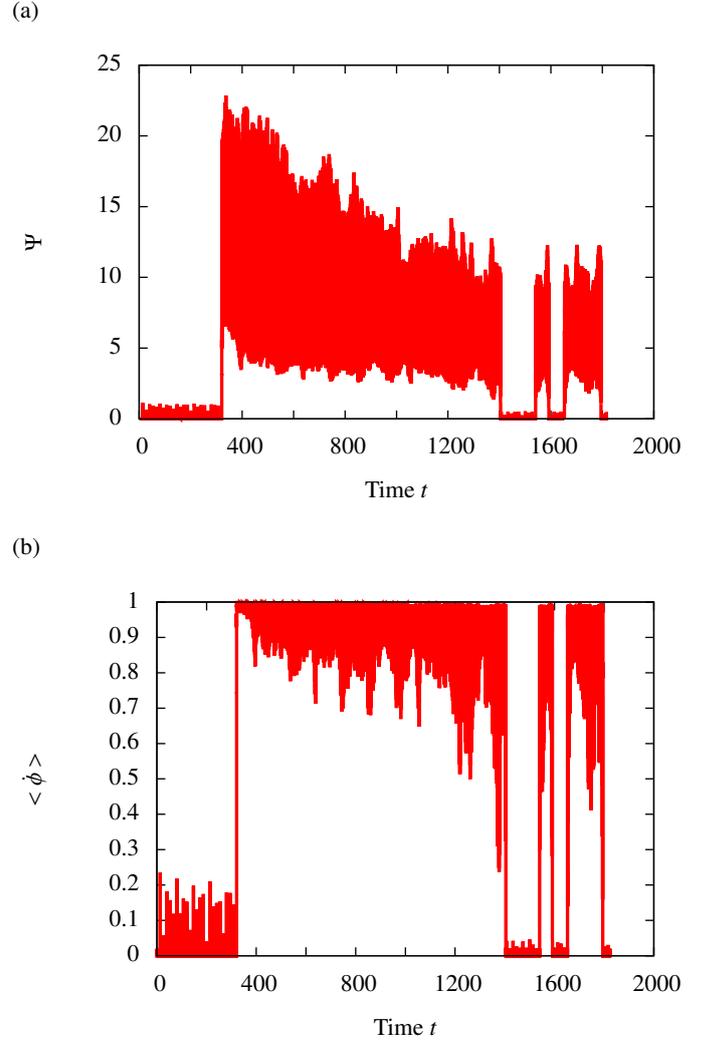

  \begin{tabular}{l}
   (a) \\
	 \scalebox{1}{\input{MeanDendriticCurrent.tex} }  \\
   (b) \\
	 \scalebox{1}{\input{Mean_Kphase.tex} }  \\   
  \end{tabular}

  \caption{ 
	(a) Mean dendritic current $\Psi(t)$ as a function of time. 
(b) Mean spiking frequency $< \dot \phi>$ as a function of time. }
  \label{fig: Mean_Current_Kphase}
\end{figure}

Therefore, in our analysis of the generalized Lighthouse model, we use
the notion of {\em states of enhanced activity}, which could be
interpreted as an epileptic seizure. A \SES is defined by a strong mean 
dendritic current 
\begin{equation}
\Psi(t)=\frac{1}{N} \sum_i \psi_i(t)
\end{equation}
A plot of the time signal of this mean current is exhibited in fig (\ref{fig: Mean_Current_Kphase}).
It is evident that periods of strong mean current and weak mean current
interchange. In order to explicitly 
determine a \SES we introduce a threshold value for the
mean current. 
A closer inspection of the time signals demonstrates that during the \SES
the spiking frequency of most of the neurons are in the saturated regime,
i.e. almost all neurons fire with the highest attainable frequency $\nu$.
This is demonstrated in fig (\ref{fig: Mean_Current_Kphase}), 
in which the mean spiking frequency $<\dot \phi>$

\begin{equation}
<\dot \phi(t)>=\frac{1}{N} \sum_i \dot \phi_i(t) 
\end{equation}

of the system is shown. 

\subsection{Initiation and Termination of States of Enhanced Activity}

\noindent States of enhanced activity emerge in the neuronal 
generalized Lighthouse model
in the following situations:
The coupling weights are initialized randomly which implies that
all to all coupling exists. One special neuron is singled out 
receiving an external signal $p_{(ext,m)}$ consisting of 
a sequence of delta peaks:
\begin{equation}
p_{(ext,m)}=\sum_n A_{p_{ext}}\delta(t-n \Delta T)
\end{equation}
The amplitude $A_{p_{ext}}$ of the peaks is chosen in such 
a way that the externally 
excited neuron always gains the maximal spiking frequency $\nu$.
The period $\Delta T$ is choosen to be $2\pi/\nu$.

The external signal mimics an external stimulation
of the network. It is well-known that epileptic seizures can be initiated by
periodic external stimuli like light flashes.
\\
\noindent As one can see in figure (\ref{fig: SLS_CurrentPhase}), 
time-intervals 
occur where all neurons emit nonvanishing dendritic currents
connected with a rapidly changing phase implying a strong spiking behaviour
of the neurons. These time intervals of strong spiking activity are interupted
by intervals where no neuron, except the externally driven one,
is active. 

States of enhanced activity are initiated by the externally driven neuron. 
Due to the external forcing this neuron spikes periodically. 
As a consequence, all neurons 
receive pulses, whose strength is proportional to the coupling weight. 
The phases of the neurons pile up until
the phase of one of the neurons reaches $2\pi$. Then, this neuron firing 
leads to an increase of the phase of the other connected neurons. 
If their phases
are close to $2\pi$, they eventually will spike and 
a chain reaction is initiated, which results in a state of enhanced activity. 

If the amplitude $A_{p_{ext}}$ is lowered, the seizure onset time increases.
Below a certain threshold, \SESs are not observed.

Termination of a \SESs is clearly due to a change of the
coupling weights $a_{km}$. During \SESs the causal nature of the 
firing of neurons, which is responsible for the built up of a network 
structure with strong couplings between causally firing neurons, 
change to acausal firing. This  eventually 
leads to a weakening of the coupling weights and, hence, to the destruction
of the state of enhanced activity. 

Let us exemplify this mechanism considering two neurons $i$ and $j$ in more
details. Assume that neuron $j$ fires before neuron $i$. As a result the 
coupling weight $a_{ij}$ will increase, whereas $a_{ji}$ decreases.
Hence, we investigate the following dynamics:
\begin{eqnarray}
\dot \phi_i &=& \Xi(\psi_i,\Theta)\nonumber \\
\dot \psi_i &=& -\gamma \psi_i +a_{ij}S_j(\phi_j)\nonumber \\
\dot \phi_j &=& \Xi(\psi_j,\Theta)\nonumber \\
\dot \psi_j &=& -\gamma \psi_j
\end{eqnarray}
We have neglected the coupling $a_{ji}\approx 0$, taking into account
that the neuron j spikes before neuron i. 

The solution for the dendritic current $\psi_i(t)$ is given in terms of
the spike-times $T_j^n$ of neuron j, which are determined by
the spiking condition $\Theta_j(T_j^n)=2\pi$:
\begin{equation}\label{strom}
\psi_i(t)=\sum_n e^{-\gamma(t-T_j^n)}a_{ij}(T_j^n)
\end{equation}
If we consider a step function $\Xi(\psi,\Theta)$, which is zero if
$\psi \le \Theta$ and $\nu$, if $\psi>\Theta$, the phase $\phi_i(t)$
changes only, if $\psi_i>\Theta$. This is the case if 
$a_{ij}(T_j^n)>\Theta$. 

Due to the acausal situation of neuron j, i.e. due to $a_{ji}\approx 0$
the dendrictic current
of neuron $j$ decays exponentially, and the spiking frequency decreases 
\begin{eqnarray}
\phi_j &=& \phi_j(0)+\int_0^t dt'\Xi(\psi_1(0)e^{-\gamma t'})
\nonumber \\
&=& \phi_j(0)+\nu ln\frac{
1+(\frac{\psi_j(0)}{A})^M}{1+(\frac{\psi_j(0)}{A})^M e^{-M\gamma t})}
\end{eqnarray}
Eventually, this leads to a stationary phase 
$\phi_j(0)+\nu ln \frac
1+(\frac{\psi_j(0)}{A})^M)$. However, for small values of the damping
constant $\gamma$, 
the dendritic current of neuron $i$ remains at a finite value, 
and, in turn its spiking frequency remains finite, provided the condition
$\sum_n e^{-\gamma(t-T_j^n)}a_{ij}(T_j^n)>\Theta$ is fullfilled.
As a consequence, the phase 
$\Phi_i$ increases more rapidly than the phase of neuron $j$: The phase of
neuron $i$ overtakes the phase of
neuron $i$ rendering the connection $a_{ij}$ acausal. This transition from 
causal to acausal firing leads to a decrease of the coupling weight
$a_{ij}$ and an increase of the coupling $a_{ji}$. 

This observation can be considered to be
consistent with the experimental findings of Schindler et al.
\cite{Schindler2008chaos} and Lehnertz et al. \cite{Lehnertz2009joneuromethod}.
These authors state that for a certain kind of epileptic seizures
synchronization and instability occur simultaneously. They argue
that the synchronization towards the end of the seizure may be
even a self-regulatory mechanism to catalyse the seizure termination. \\

\noindent 
The described behaviour exists for a certain range of values of the damping
constant $\gamma$, $\gamma_C^l<\gamma < \gamma_c^u$. 
This has been investigated by a numerical investigation
of the generalized Lighthouse model. This range depends on the system size.

The described mechanism only works, if the damping constant $\gamma$ of the 
dendritic current is large enough. Therefore 
a lower critical damping constant $\gamma_c^l$ exists. 
On the opposite, the chain reaction can not start, if the damping constant 
exceeds a certain critical upper 
value $\gamma_c^u$. The neurons will start to spike, but the dendritic 
current of 
the neuron is decaying too fast and therefore the phase velocity is not high 
enough to preserve the firing and initialize a state of enhanced activity.

\subsection{Analysis of Statistical Behaviour}

\begin{table}[t]
\centering
\begin{tabular}{l l}
\hline
\hline
Numerical parameters: & Naka-Rushton relation: \\
\hline
Time step $dt$ = $0.01$ &  maximal spiking rate $\nu$: $1.0$ \\
&Threshold $\theta$: $10.0$ \\
&Steepness $M$: $3$ \\
\hline
\hline
External signal: & Neuronal network: \\
\hline
Amplitude $A_{p_{ext}}$: $10.0$ & Number Neurons $N$: $50$ \\
Forcing Periodic $T_{p_{ext}}$: $1.0$  &  Damping constant $\gamma$: $0.7$ \\
 & Enhancement constant $c_m$: $5.0$ \\
\hline
\hline
\multicolumn{2}{l}{Learning parameters:} \\
\hline
Potentiation constant $\Delta$: $1.0$ & Depression constant $r$: $1.0$ \\
Potentiation width $\tau_A$: $0.2$ & Depression width $\tau_B$: $0.2$ \\
Recovery rate $\tau_{r \sigma}$: $10.0$ & Fatigue rate $\tau_{l \sigma}$: $10.0$ \\
Chemical fraction $u_A$: $0.9$ & Chemical fraction $u_B$: $0.9$ \\
\hline
\hline
\end{tabular}

\caption{Numerical coefficients used in the simulations, 
are shown.}
\label{table: Numerical coefficients}
\end{table}

\noindent In the following, we present results of the 
investigation of the statistical behaviour of the 
states of enhanced activity. In order to create sufficient 
statistics long time simulations of the generalized Lighthouse model
and measurements of the mean dendritic currents have been performed. 
The used numerical parameters are shown in table 
(\ref{table: Numerical coefficients}).
$27$ runs have been realized, each with $47 \cdot 10^7$ timesteps. 
In total $10,000$ events, which could be classified as \SES, have been 
measured.
The duration of a state of enhanced activity, the interevent waiting times between two subsequent
states of enhanced activity, and the total energy of a state of enhanced energy have been calculated, 
The energy of a state of enhanced activity is defined to be 
proportional to the sum of the squares of all dendritic currents:
\begin{equation}
E= R \left[ \frac{1}{N} \sum_{i} \psi_i(t)^2 \right]
\end{equation}
Based on this data, 
we have calculated the probability density functions (PDF) of the energy, 
the event duration and the interevent waiting time.
Furthermore the Omori- and inverse-Omori-Law was determined in the same way 
as described by Osorio et al. \cite{Osorio2010pre}. Furthermore,
the conditional expected waiting-time was calculated. 

\begin{figure}[t]
	\input{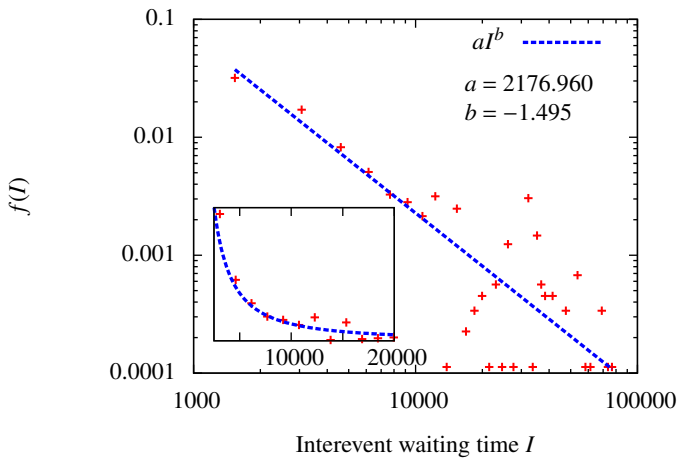}
  \caption{The logarithm of the probability density function $f(I)$ of 
the interevent-intervals of the states of enhanced activity as a function of $ln I$. 
	  The inset shows the same data points in $f(I)$-$I$ coordinates . 
A power law is fitted to the data.}
  \label{fig: PDF_InterEvent}
\end{figure}

\noindent In order to assess the scaling behaviour 
of the various quantities like energy of an state of enhanced activity, duration of an state of enhanced activity, or
the interevent times between two subsequent states of enhanced activity, 
the scaling coefficients of the various quantities are calculated by the 
method introduced by Newman \cite{Newman2006arxiv}.
It is based on a maximum likelihood estimation of the scaling coefficient,
The estimation is only applied in intervals $[x_{min}, x_{max}]$ of the data in 
which the power law behaviour holds.

\subsection{Interevent Waiting Time}

\noindent As stated by Osorio et al and Saichev et al \cite{Osorio2010pre, Saichev2006Jgr}, the \PDF
of the interevent waiting time for earthquakes is described by a power law 
with a scaling
coefficient of $-1.1$. Osorio et al \cite{Osorio2010pre} calculated the PDF 
for data sets of 
epileptic seizures. They obtained  
an approximate value of $-\frac{3}{2}$ for the scaling coefficient.

From our numerical computations of the generalized Lighthouse model
we extract a scaling coefficient of $-1.495 \pm 0.149$ for $[x_{min} = 100, x_{max} = 18000]$, which
perfectly agrees with the analysis of Osorio \cite{Osorio2010pre}.
The data is exhibited in figure (\ref{fig: PDF_InterEvent}).
It is obvious that the power law fits quite well the measured behaviour
for small and medium interevent waiting times. For larger values of the 
interevent times the data points are scattered around the fitted power law. 
This could be explained, following  
Osorio et al \cite{Osorio2010pre}, by the assumption that 
the probability density function exhibits different power law regimes, 
a possibility which has first been pointed out by \cite{Saichev2006Jgr}. 
However, it could also be attributed to the restricted sample size.
We remind the reader that the behaviour in the
regime of large interevent times is of considerable interest with respect to the analysis of extreme events in complex systems. 

\subsection{Gutenberg-Richter-Law}

\begin{figure}[t]
    \input{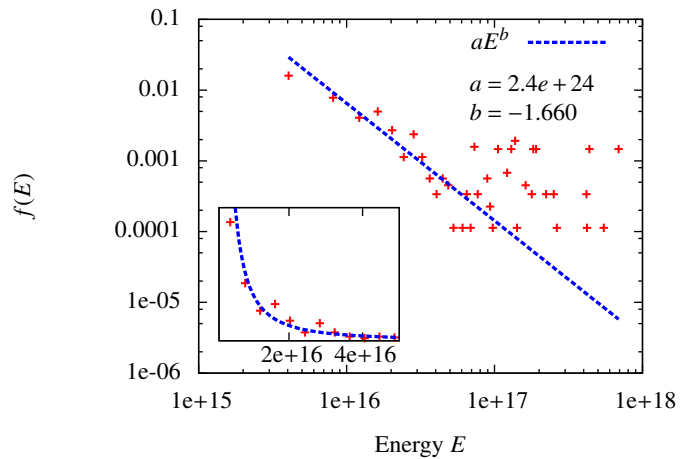}
    \caption{The probability density function $f(E)$ of the total energy of state of enhanced activity is shown. The coefficient is $-1.660 \pm 0.220$ in the interval
    $[3.7 \cdot 10^{15}, 4 \cdot 10^{16}]$.}
      \label{fig: PDF_Energy}
\end{figure}

\noindent The Gutenberg-Richter-law states that the \PDF for the earthquake seismic moment $S$ 
exhibits a power law distribution. Similarly, the analysis for
epileptic seizures
by \cite{Osorio2010pre} yields a value close to $-\nicefrac{5}{3}$. 
Furthermore they showed that
a power law with the same scaling exponent is obtained 
for the energy distribution of epileptic seizures.
The results of our calculations, which are summarized
in figure (\ref{fig: PDF_Energy}), support this assumption.

\begin{figure}[t]
	\begin{flushleft}
	\input{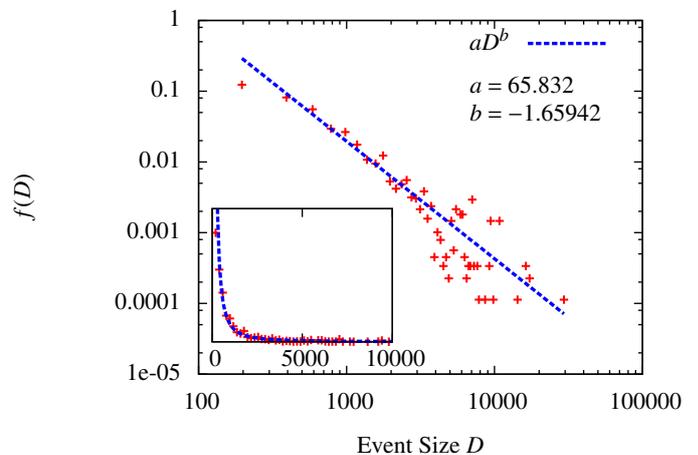}
	\end{flushleft}
	\caption{The logarithm of the 
probability density function $f(D)$ of the duration as a function of log D. 
The power law fit yields
	    a value for the scaling coefficient of $-1.659 \pm 0.155$ in the interval $[200, 3783]$. The inset exhibits $f(D)$ as a function of $D$.}
	  \label{fig: PDF_EventSize}
\end{figure}

\noindent  As for the statistics of the interevent times
the behaviour of the PDF deviates from the powerlaw for high values of $E$, 
which can be explained by the same reasonings as discussed above.
If one considers the PDF of the duration of the states of enhanced activity, 
which is plotted in figure (\ref{fig: PDF_EventSize}), 
one can see that both PDFs have a similar scaling exponent. 

\begin{figure}[h]
\begin{center}
    \input{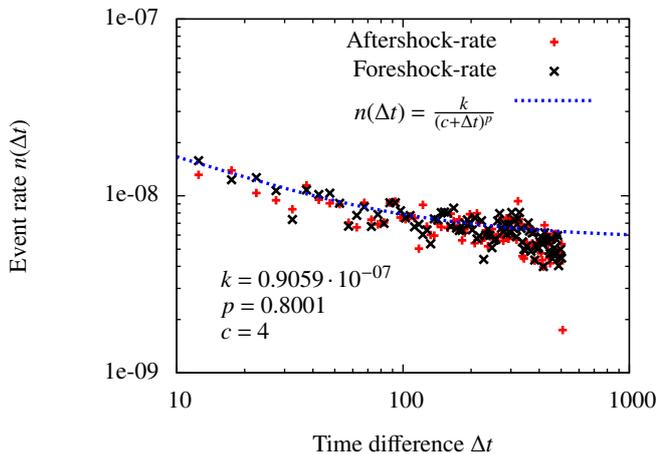}   
  \caption{The foreshock and aftershock rates for states of enhanced activity are shown in a double logarithmic scale. The time axis for the foreshock rate
  is inverted. The Omori-Law is fitted into the data points and the fitting-parameters are shown.}
  \label{fig: OmoriLaw}
\end{center}  
\end{figure}

\noindent Therefore, a direct relationship between energy $E$ and duration
of the \SES should exist. This relationship can be seen on the basis of
the definition
\begin{align}
	E(S)= \frac{R}{M}\sum_{t_a,t_e} \psi_m(t_n)^2 = <I> \cdot D(S).
\label{eqn: Energy Duration}
\end{align}
of the energy contained in the \SES between the time instants
$t_a$, $t_e$. If we replace approximately the quantity
$\psi_m(t_n)^2$ by a mean intensity
$<I>$ we obtain a direct relationship between $E$ and duration $D$, and, hence,
the same scaling behaviour.

Assuming the validity of this relation, one can conclude that 
all states, independent of their duration, have the same static distribution of the intensity $<I>$ and
therefore the same distribution of the dendritic current $\Psi$. 

\subsection{Omori Law and Inverse Omori Law}

Earthquakes are rare events, which are accompanied by 
the occurence of aftershock and foreshock events. The rate $n(\Delta t)$ of 
these events follows the so-called 
Omori- and inverse Omori-Law as a function  of the time distance $\Delta t$
from the main event:
\begin{align}
log(n(\Delta t)) = \log(k) - p ~ \log(c+\Delta t)
\label{eqn: Log Omori-Law}
\end{align}
Both laws are  empirical. These laws originally were 
introduced by Omori et al \cite{Omori1894JCS} and 
expanded by Utsu et al \cite{Utsu1995JPE}.

Osorio et al \cite{Osorio2010pre} pointed out
that the Omori- and inverse Omori-Law can also be formulated for 
epileptic seizures. The analysis of data generated by the generalized Lighthouse model supports the validity of these laws. Our results are presented
in figure (\ref{fig: OmoriLaw}). The Omori-Law is fitted to the data. 
Both, the foreshock and aftershock rates can be represented by
the same fit. This implies that the shock rates 
are not only described by the Omori-Law, but are symmetric. 
The results are consistent with the results for
data of epileptic seizures \cite{Osorio2010pre} and are similar 
to the earthquake data \cite{Osorio2010pre,Shcherbakov2004gpl,Christensen2002pnasu,Bak2002prl}.  

\subsection{Expected Waiting Time}

\begin{figure}[t]
\begin{center}
  \input{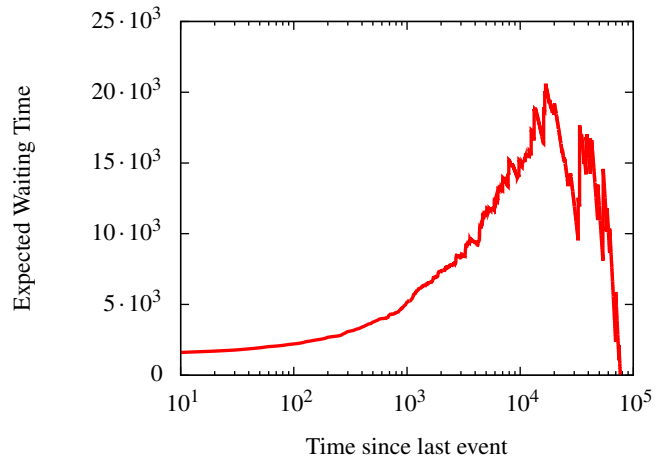}   
  \caption{Averaged waiting time to the next seizure-like-state 
as a function of the time distance to the last event. }
  \label{fig: WaitingTime}
\end{center}
\end{figure}

An interesting question is how the waiting time, which is the time 
to the next event, depends on the time that has already passed 
since the last 
event. The results of our numerical simulations are shown in figure 
(\ref{fig: WaitingTime}).

As one can see, the expected waiting time to the next event increases 
as a function of
the distance to the last event 
to a maximum. Then the expected waiting time rapidly drops down to zero.
The intersection with the x-axis marks the maximal time that can pass 
before the next state occurs. This is clearly a finite size effect and 
depends on the number of neurons of the model.
A similar behavior is observed for epileptic seizures as well as
earthquakes \cite{Osorio2010pre}. 

\section{Summary and Outlook}

\noindent We combined the Lighthouse model introduced by Haken 
\cite{Haken2006book} with a STDP-learning algorithm 
to a model in which states of enhanced activity emerge.
It has been shown that these states share characteristic features reported 
for time signals of epileptic seizures. Defining states of enhanced activity
via the energy $E$ we demonstrated the validity of the
Gutenberg-Richter law, the Omori
and inverse Omori laws and showed similarities in
the statistics of inter-seizure intervals.
We want to emphasize that our model yields scaling exponents rather
close to the ones reported in the work of Osorio et al. 
\cite{Osorio2010pre}. Therefore, a detailed investigation of
the generalized Lighthouse model should allow for an analytical assessment of
these scaling exponents. Furthermore, we have  
described the mechanism which leads to a termination of
the states of enhanced activity as a transition from
causal to acausal firing, connected with an immediate restructuring of
the network topology due to STDP.   
Due to the closeness to experimental findings it would be interesting
to explore the phase diagram of this generalized Lighthouse
model in more details and analyze experimental data, whether the mechanism
of termination of \SES can also be detected for epileptic seizures. For the 
future, we plan to
perform an extended analysis in order to
assess the transitions and bifurcation scenarios 
leading to the regime, where \SES arise, from  
ordered or disordered states adding dynamical details to the 
schematic phase diagram given in Osorio \cite{Osorio2010pre} locating seizure
states in a control parameter space spanned by the variables termed
{\em interaction strength} and {\em heterogeneity}. 
Additionally, we plan to extend our
analysis to larger data samples extending our numerical calculations to 
longer time periods. This will allow us to resolve the PDF's of extreme events, along the lines suggested by Sornette \cite{Sornette2010IJOTSE}.
In this way we want to contribute to the 
recent discussions on the existence of dragon kings
\cite{Sornette2010IJOTSE, Sornette-review} and 
the possibility of predictions
of extreme events, recently reviewed in \cite{Physical-Journal}.

\bibliography{Draft}{}

\end{document}